# Single Vanadium Dioxide Nanoparticle-Enabled Plasmonic switch with Thermal and Electronic Reconfigurability

Gregory Tanyi
Department of Electrical and Electronic Engineering, Faculty of Engineering and Information Technology, The University of Melbourne, Parkville VIC 3010, Australia &
Sustainable Mining Technologies, Commonwealth Scientific Industry and Research Organiasation,
1 Technology Court Pullenvale,
Email: gtanyi@student.unimelb.edu.au, gregory.tanyi@csiro.au

Daniel Peace
Australian Research Council Centre of Excellence for Engineered Quantum Systems & School of Mathematics and Physics, The University of Queensland, St Lucia 4067, Queensland, Australia.
d.peace@uq.edu.au

Elliot Cheng
1 Centre for Microscopy and Microanalysis, St Lucia 4067, Queensland, Australia.
h.cheng6@uq.edu.au

Mohammed Taha
Department of Electrical and Electronic Engineering, Faculty of Engineering and Information Technology, The University of Melbourne, Parkville VIC 3010, Australia
mohammad.taha@unimelb.edu.au

Guanghui Ren
Integrated Photonics and Applications Centre, School of Engineering, RMIT University, Melbourne, Victoria, Australia
guanghui.ren@rmit.edu.au

Xuan Hiep Dinh
Integrated Photonics and Applications Centre, School of Engineering, RMIT University, Melbourne, Victoria, Australia
S3923133@student.rmit.edu.au

Arnan Mitchell
Integrated Photonics and Applications Centre, School of Engineering, RMIT University, Melbourne, Victoria, Australia
arnan.mitchell@rmit.edu.au

Karsten Hoehn
Sustainable Mining Technologies
Commonwealth Scientific Industry and Research Organiasation,
1 Technology Court Pullenvale,
karsten.hoehn@csiro.au

Christina Lim
Department of Electrical and Electronic Engineering, Faculty of Engineering and Information Technology, The University of Melbourne, Parkville VIC 3010, Australia
chrislim@unimelb.edu.au

Ranjith Unnithan
Department of Electrical and Electronic Engineering, Faculty of Engineering and Information Technology, The University of Melbourne, Parkville VIC 3010, Australia
r.ranjith@unimelb.edu.au

*Abstract*—We present an integrated switch that combines plasmonic and neuromorphic technologies with a single sub-stoichiometric $VO_{2-x}$ nanoparticle. The presented device acts as a versatile plasmonic switch with dual thermal and electrical reconfigurability leveraging the near-room temperature phase transition of the $VO_{2-x}$ nanoparticles combined with the rapid phase recovery to drive the device. The change in both the optical and electrical properties of the $VO_{2-x}$ nanoparticle enables simultaneous optical and electrical readouts making the plasmonic device suitable as a phase change memory cell which is crucial in the convergence of computing and communication technologies. Our demonstration of reversible electrical switching, evidenced by a 6dB modulation depth and concurrent optical and electrical outputs, signifies a major stride in merging electronic and photonic functionalities within phase-change material devices. This novel strategy not only harmonizes optical communication with electronic computing but also advances the development of sophisticated integrated neuromorphic devices.

## I. INTRODUCTION

Recent advances in brain-inspired computing have opened new avenues in scientific research, with machine learning algorithms demonstrating remarkable capabilities in areas traditionally dominated by human intelligence, such as speech and pattern recognition. Despite significant progress in neuromorphic software, hardware development has lagged, primarily due to the von Neumann bottleneck, which arises from the separation of processing and storage units. This bottleneck has catalyzed the search for efficient hardware solutions capable of integrating storage and processing functions, showing promise in applications like integrated photonic switches and optical neural networks[1]-[4].

To address the gap in neuromorphic hardware, researchers have developed devices that merge storage and processing capabilities. Phase-change materials (PCMs) are particularly promising for these applications, as they can switch between states with distinct optical and electrical properties, enabling simultaneous storage and logic operations. Integrated phase-change optical memory systems combine PCMs with optical technology for efficient, high-density data storage[5].

However, integrating phase-change materials for both photonic and electrical applications poses challenges due to the conflicting electrical and optical requirements as electrical switching of PCMs requires extremely small separation between the electrodes to generate a strong electric field while the optical response is diffraction limited. Vanadium dioxide (VO$_2$), a Mott material, has been explored for its potential in neuromorphic and photonic switching applications, owing to its large changes in refractive index and resistivity during phase transitions [6]-**Error! Reference source not found.**. Sub-stoichiometric VO$_{2-x}$ nanoparticles, with their reduced transition temperature and rapid phase transitions, are emerging as candidates for reconfigurable photonics switches and computing devices [6].

Surface plasmon polaritons (SPPs) enable the focusing of light beyond the diffraction limit making plasmonic devices appealing for subwavelength photonic applications. In this work, by combining plasmonics with silicon photonics and phase change materials, we present a versatile switch that operates across a wide wavelength range (1500 – 1560nm), and a novel phase-change memory cell that can be accessed both electrically and optically.

## II. MATERIALS AND METHODS

### A. Device design

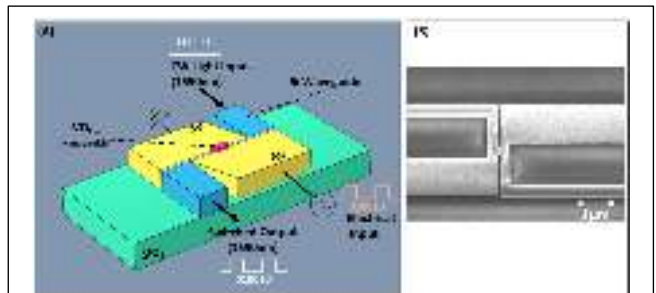

Figure 1 (a) Three-dimensional illustration of the plasmonic device driven by a single vanadium oxide nanoparticle highlighting the presence of silicon waveguides, a gap plasmon waveguide with a metal-insulator-metal structure, and the incorporation of a vanadium oxide nanoparticle within the plasmonic waveguide. The device switches states by switching the phase of the VO$_{2-x}$ nanoparticle which exhibits a large change in refractive index and resistance as it switches from an insulating to a metallic state, a process activated by both heat and electrical stimuli. (b) SEM image of the fabricated device showing compact plasmonic slot with a single VO$_{2-x}$ nanoparticle after device fabrication.

The presented plasmonic device, illustrated in Figure 1a, utilizes a hybrid orthogonal junction plasmonic coupling scheme, featuring a metal-insulator-metal gap with a 90nm separation, orthogonal to 450nm-wide silicon waveguides and 150nm-thick gold electrodes. Central to the device is a vanadium oxide nanoparticle positioned within the plasmonic slot, acting as the active region.

Light coupling from a single-mode fiber occurs via a grating coupler, with photons at the initial orthogonal junction being coupled to surface plasmon polaritons that traverse the plasmonic waveguide. The device is simulated using the finite element method implemented in COMSOL Multiphysics [9-12] and fabricated on a commercial SOI wafer with a 220nm device Silicon layer.

The device operates in two states: OFF, with the VO2-x nanoparticle in its semiconductor phase, allowing surface plasmon polaritons to travel through the plasmonic waveguide; and ON, where an electric field or thermal energy induces a transition to the metallic phase, attenuating the surface plasmon polaritons. This transition results in modulated optical transmission levels due to the change in nanoparticle phase, effectively altering the device's optical output. The phase transition of the nanoparticles also leads to a change in the resistivity of the nanoparticles which alters the devices' electrical output.

## III. RESULTS AND DISCUSSION

We demonstrate both thermal and electronic reconfigurability of the plasmonic device. Figure 2a illustrates the thermal switching of the device enabled by the thermally trigger first order insulator-to-metal transition of the vanadium dioxide (VO$_{2-x}$) nanoparticle. The device exhibits lower optical transmission in its ON state, where the nanoparticle is in the metallic phase, as compared to the OFF state, with the nanoparticle in the semiconductor phase. This difference in optical transmission, quantified as the extinction ratio, reaches a minimum of 7dB over a wavelength range of 1500nm to 1600nm. The operational wavelength range is primarily focused on the C band. In addition to thermal reconfigurability, the device also demonstrates reversible electrical switching, with the capability for both optical and electrical readouts.

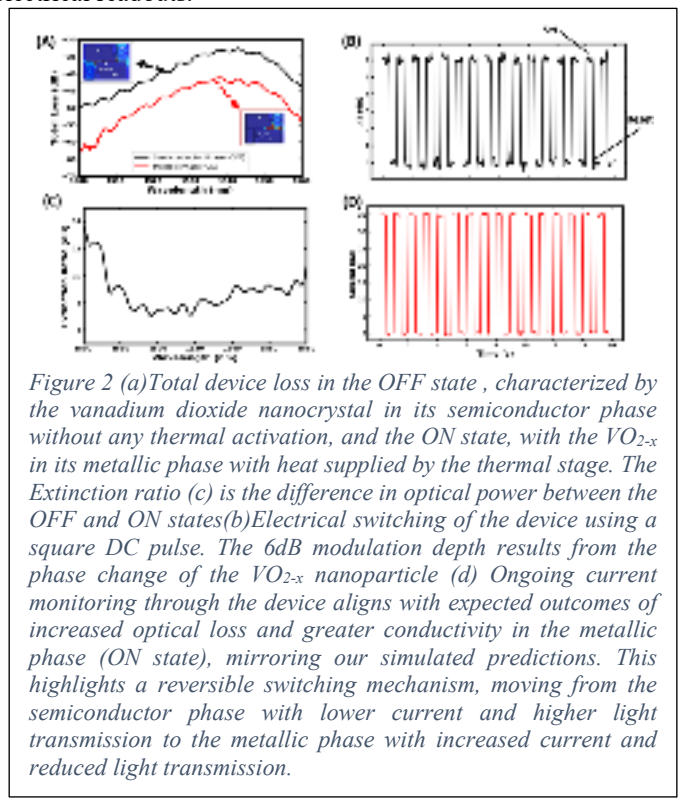

Figure 2 (a)Total device loss in the OFF state, characterized by the vanadium dioxide nanocrystal in its semiconductor phase without any thermal activation, and the ON state, with the VO$_{2-x}$ in its metallic phase with heat supplied by the thermal stage. The Extinction ratio (c) is the difference in optical power between the OFF and ON states(b)Electrical switching of the device using a square DC pulse. The 6dB modulation depth results from the phase change of the VO$_{2-x}$ nanoparticle (d) Ongoing current monitoring through the device aligns with expected outcomes of increased optical loss and greater conductivity in the metallic phase (ON state), mirroring our simulated predictions. This highlights a reversible switching mechanism, moving from the semiconductor phase with lower current and higher light transmission to the metallic phase with increased current and reduced light transmission.

The switching mechanism is powered by a relatively low energy requirement of approximately 0.712 fJ. Through the application of a continuous wave laser and subsequent monitoring via a low noise photodetector, the device's optical output can be continuously assessed. Electrical phase changes are initiated through rectangular DC signals (3.5V set and 1.9V reset), enabling the VO2-x nanoparticle to transition between semiconductor and metallic phases, significantly affecting both the current flow and optical transmission. Notably, this electrical reconfigurability results in a distinct 6dB switching ratio, achieved using a single VO2-x nanocrystal. This dual reconfigurability, both thermal and electrical, underscores the device's versatility and potential in advancing plasmonic neuromorphic computing technologies.

## IV. CONCLUSION

We demonstrate an innovative plasmonic device that exhibits thermal and electronic reconfigurability with simultaneous optical and electrical readouts offering an approach to the integration of computing and communication technologies. The wide operational wavelength range and scalable manufacturing of the device mean that integrating vanadium oxide nanoparticles into plasmonic slots could lead to the development of ultra-compact, on-chip plasmonic switches and optoelectronic neuromorphic computing devices with built-in memory capabilities.

## ACKNOWLEDGMENT

This work was performed in part at the Melbourne Centre for Nanofabrication (MCN) in the Victorian Node of the Australian National Fabrication Facility (ANFF).

## REFERENCES

[1] J. Lappalainen, J. Mizsei, and M. Huotari, "Neuromorphic thermal-electric circuits based on phase-change VO2 thin-film memristor elements," *Journal of Applied Physics*, vol. 125, no. 4, 2019. doi:10.1063/1.5037990

[2] J. von Neumann, *The Computer and the Brain*. New Haven: Yale University, 1958.

[3] A. Sebastian, M. Le Gallo, R. Khaddam-Aljameh, and E. Eleftheriou, "Memory devices and applications for in-memory computing," *Nature Nanotechnology*, vol. 15, no. 7, pp. 529–544, 2020. doi:10.1038/s41565-020-0655-z

[4] A. Sebastian, M. Le Gallo, and E. Eleftheriou, "Computational phase-change memory: Beyond von neumann computing," *Journal of Physics D: Applied Physics*, vol. 52, no. 44, p. 443002, 2019. doi:10.1088/1361-6463/ab37b6

[5] Y.-Y. Au, H. Bhaskaran, and C. D. Wright, "Phase-change devices for simultaneous optical-electrical applications," *Scientific Reports*, vol. 7, no. 1, 2017. doi:10.1038/s41598-017-10425-8

[6] G. Tanyi, C. Lim, and R. R. Unnithan, "Design of an on-chip vanadium dioxide driven plasmonic modulator based on hybrid orthogonal junctions on silicon-on-insulator," *CLEO 2023*, 2023. doi:10.1364/cleo_fs.2023.fw4c.3

[7] Rajasekharan, R et al, "Nanophotonic Three-Dimensional Microscope," Nano Lett. 2011, 11, 7, 2770–2773 -1 (7), 2011.

[8] M. Taha *et al.*, "Infrared modulation *via* near-room-temperature phase transitions of vanadium oxides & Core–Shell Composites," *Journal of Materials Chemistry A*, vol. 11, no. 14, pp. 7629–7638, 2023. doi:10.1039/d2ta09753b

[9] M. Sun, et al., "Design of Plasmonic Modulators with Vanadium Dioxide on Silicon-On-Insulator, IEEE Photonics Journal," vol. 9, no. 3, pp. 1-10, 2017

[10] "Comsol Multiphysics," *https://www.comsol.com/*.

[11] Q. Dai et al., "Transparent liquid-crystal-based microlens array using vertically aligned carbon nanofiber electrodes on quartz substrates," Nanotechnology 22 (11), 115201, 2011